# One-way Optomechanical Interaction between Nanoparticles

**Amir M. Jazayeri[*], Sohila Abdelhafiz, and Aristide Dogariu[†]**
*The College of Optics and Photonics (CREOL), University of Central Florida, Orlando, FL, USA*
[*]*amir@ucf.edu*, [†]*adogariu@creol.ucf.edu*

**Abstract**- Within a closed system, physical interactions are reciprocal. However, the effective interaction between two entities of an open system may not obey reciprocity. Here, we describe a non-reciprocal interaction between nanoparticles which is one-way, almost insensitive to the interparticle distance, and scalable to many particles. The interaction we propose is based on the non-conservative optical forces between two nanoparticles with highly directional scattering patterns. However, we elucidate that scattering patterns can in general be very misleading about the interparticle forces. We introduce zeroth- and first-order non-reciprocity factors to precisely quantify the merits of any optomechanical interaction between nanoparticles. Our proposed one-way interaction could constitute an important step in the realization of mesoscopic heat pumps and refrigerators, the study of non-equilibrium systems, and the simulation of non-Hermitian quantum models.

## I. INTRODUCTION

A physical 'system' is defined by the observer, and it is never closed. Rather, it interacts with one or several thermal and/or electromagnetic baths [1-3]. In steady state, if such a state exists, the system may or not be in thermal equilibrium. The most intuitive example of a non-equilibrium steady state is the case of a mechanical oscillator in contact with two thermal baths of different temperatures [4]. Another example is the steady state of a mechanical oscillator which interacts with a thermal bath of a certain temperature while also driven by an optical force with non-symmetric spectral density [5-9]. It should be noted that in both examples, the interaction between the mechanical oscillator and its outside world is 'reciprocal', and therefore, the fluctuation-dissipation theorem in its original form [10] holds.

A question arises now. Can the 'effective' interaction between two entities 'A' and 'B' be non-reciprocal? The term 'effective' refers to the fact that such an interaction needs to be mediated by another entity 'C' excluded from the system that includes 'A' and 'B'. Maybe the most well-known example of non-reciprocity is the Faraday effect. Another example very similar to the Faraday effect is non-reciprocal light transmission through a nano-waveguide placed in the vicinity of a quantum dot [11], where the quantum dot plays the role of the mediating entity 'C'. As a different example, the entities 'A' and 'B' can be two quantum dots while the mediating entity 'C' is the set of the optical modes of a nano-waveguide [12]. Non-reciprocal interactions are also ubiquitous in active and living matter [13-16]. Non-reciprocal interactions have also been realized using active components in electronic circuits [17] or active control on robotic elements of mechanical systems [18].

In this article, we seek a non-reciprocal interaction between two dielectric nanoparticles 'A' and 'B', where the mediating entity 'C' is the set of the electromagnetic modes in free space (or the homogenous medium surrounding the nanoparticles). As is depicted in Fig. 1, the interaction between 'A' and 'B' is defined based on the exerted optical force $\vec{f}_A$ on 'A' due to the presence of 'B' together with the exerted optical force $\vec{f}_B$ on 'B' due to the presence of 'A'. Furthermore, the interaction we seek is to be (i) nearly one-way irrespective of the interparticle distance, and

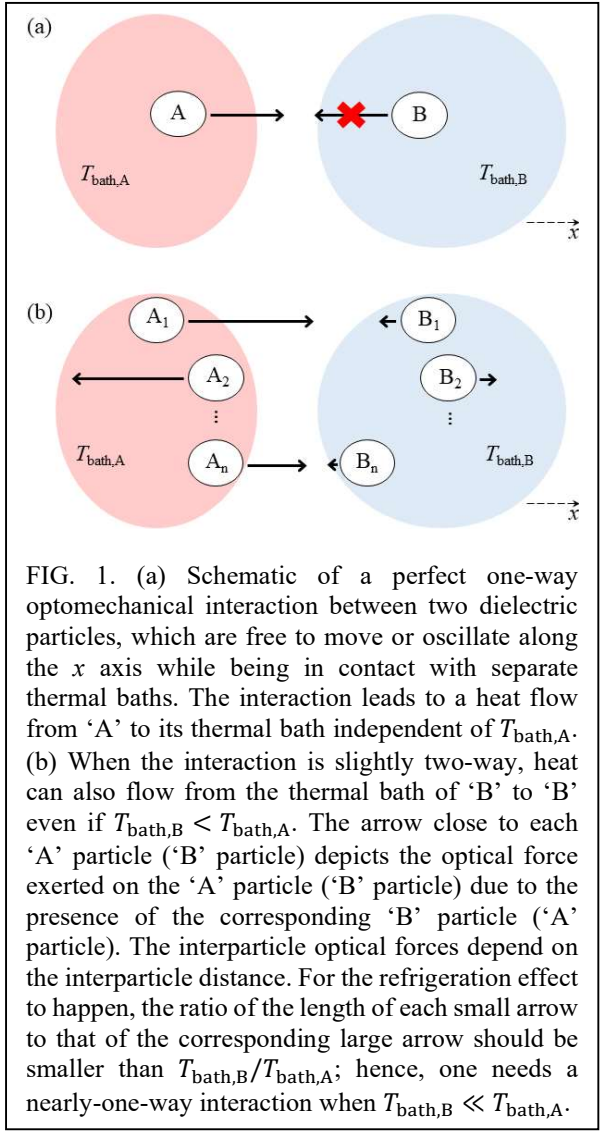

FIG. 1. (a) Schematic of a perfect one-way optomechanical interaction between two dielectric particles, which are free to move or oscillate along the $x$ axis while being in contact with separate thermal baths. The interaction leads to a heat flow from 'A' to its thermal bath independent of $T_{\text{bath,A}}$. (b) When the interaction is slightly two-way, heat can also flow from the thermal bath of 'B' to 'B' even if $T_{\text{bath,B}} < T_{\text{bath,A}}$. The arrow close to each 'A' particle ('B' particle) depicts the optical force exerted on the 'A' particle ('B' particle) due to the presence of the corresponding 'B' particle ('A' particle). The interparticle optical forces depend on the interparticle distance. For the refrigeration effect to happen, the ratio of the length of each small arrow to that of the corresponding large arrow should be smaller than $T_{\text{bath,B}}/T_{\text{bath,A}}$; hence, one needs a nearly-one-way interaction when $T_{\text{bath,B}} \ll T_{\text{bath,A}}$.

(ii) scalable to many particles. The first condition means that the magnitude of $\vec{f}_B$ is to be much smaller than the magnitude of $\vec{f}_A$ irrespective of the interparticle distance, although $\vec{f}_A$ itself depends on the interparticle distance.

Non-conservativeness of the optical forces has a decisive role in the one-way interaction we propose. This is in stark contrast to any optomechanical system that builds an effective non-reciprocal interaction between two mechanical modes by using conservative optical forces. In such systems [19], the optical mode of a highly engineered cavity is coupled to two of its mechanical modes via conservative optical forces. As a result, if one ignores the unwanted decay rates of the optical and mechanical modes, the effective interaction Hamiltonian between the mechanical modes (after eliminating the optical mode) is Hermitian.

As depicted in Fig. 1(a), a perfect one-way interaction, with a zero $\vec{f}_B$ and an $\vec{f}_A$ highly dependent on the interparticle distance, leads to a maximally directional information flow from 'B' to 'A' [20]. Such an interaction also leads to a heat flow, independent of the temperature $T_{\text{bath,A}}$ of the thermal bath of 'A', from 'A' to its thermal bath [20]. More interestingly, as depicted in Fig. 1(b), by making the interaction slightly two-way but still highly non-reciprocal (i.e., having a small non-zero $\vec{f}_B$ with a direction opposite to the direction of $\vec{f}_A$), heat can also flow from the thermal bath of 'B' to 'B' even when $T_{\text{bath,B}} < T_{\text{bath,A}}$ [20,21]. This refrigeration effect needs $|\vec{f}_B|/|\vec{f}_A|$ to be smaller than $T_{\text{bath,B}}/T_{\text{bath,A}}$; hence, one needs a nearly-one-way interaction when $T_{\text{bath,B}} \ll T_{\text{bath,A}}$. By increasing the number of the interacting pairs 'A' and 'B', one can expect that the particles will eventually affect the temperatures of their corresponding baths, or, in other words, the baths will not be actual baths anymore. At mesoscopic scales, such a scheme can in fact lead to new types of heat pumps and refrigerators that operate based on a mechanism different than those examined recently [22-25].

## II. NON-RECIPROCITY FACTORS

Before going into the details of our proposal, let us precisely define and quantify the level of non-reciprocity in an optomechanical interaction between two particles 'A' and 'B'. With respect to a certain action-reaction axis $\hat{\alpha}$, the interaction is reciprocal if and only if $\hat{\alpha} \cdot (\vec{f}_A + \vec{f}_B)$ is zero.

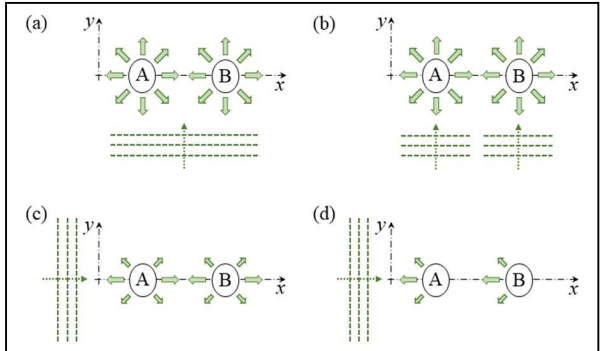

FIG. 2. Different systems of two identical nanoparticles. The thick green arrows illustrate the far-field scattering pattern of each particle in the absence of the other particle. The external electric fields in (a,b) are $z$-polarized. The external electric fields in (c,d) can have any linear polarizations parallel to the $yz$ plane, but the $xy$ scattering patterns in the schematic are drawn for $y$-polarized external electric fields. In (b), the particles are illuminated by waves that have an engineered relative phase difference at the $xz$ plane. Each particle in (a-c) can be modeled by a polarizable electric dipole, whereas each particle in (d) can be modeled by the combination of a polarizable electric dipole and a polarizable magnetic dipole.

Here, $\vec{f}_A$ reads $\vec{F}_A - \vec{F}_{A,\text{iso}}$ in terms of the forces $\vec{F}_A$ and $\vec{F}_{A,\text{iso}}$ that denote the exerted optical forces on 'A' in the presence and absence of 'B', respectively. A similar definition applies to the exerted optical forces on 'B'. Importantly, this reciprocity condition should apply irrespective of the particles' positions $\vec{r}_A$ and $\vec{r}_B$. However, in practice, this is not the case for the optomechanical interactions, and the notion of reciprocity must be carefully examined. Here we introduce a general way to quantify an optomechanical non-reciprocity.

We define the interaction to be reciprocal in the *zeroth order* with respect to a specific direction $\hat{\alpha}$ if and only if $\hat{\alpha} \cdot (\vec{f}_A + \vec{f}_B)$ is zero for given positions $\vec{r}_A$ and $\vec{r}_B$. Thus, the zeroth-order non-reciprocity factor $NRF_{\hat{\alpha}}^{(0)}$ with respect to $\hat{\alpha}$ can be defined as

$$\frac{|\hat{\alpha}\cdot(\vec{f}_A+\vec{f}_B)|}{|\hat{\alpha}\cdot(\vec{f}_A-\vec{f}_B)|+|\hat{\alpha}\cdot(\vec{f}_A+\vec{f}_B)|}, \quad (1)$$

which is never negative or larger than unity. This factor is equal to zero, one-half, and unity in the case of a fully reciprocal interaction, one-way interaction, and fully-non-reciprocal interaction with respect to $\hat{\alpha}$.

The value of $NRF_{\hat{\alpha}}^{(0)}$ does not yield any information about the dynamic case, where, for instance, the particles are free to move around certain locations. To overcome this issue, we define that an optomechanical interaction is reciprocal in the *first order* with respect to two specific directions $\hat{\alpha}$ and $\hat{\beta}$, which may or may not be different, if and only if the conditions $\hat{v} \cdot \partial(\vec{f}_A + \vec{f}_B)/\partial w_P = 0$ are all met. Here 'P' is either 'A' or 'B', and each of $\hat{v}$ and $\hat{w}$ is either $\hat{\alpha}$ or $\hat{\beta}$. Also, $w_P$ denotes $\hat{w} \cdot \vec{r}_P$, and the derivatives are calculated at given locations $\vec{r}_A$ and $\vec{r}_B$. We can now define a set of first-order non-reciprocity factors $\{NRF_{\hat{v},w_P}^{(1)}\}$ with respect to the directions $\hat{\alpha}$ and $\hat{\beta}$ as

$$\left\{ \frac{\left|\hat{v}\cdot\frac{\partial(\vec{f}_A+\vec{f}_B)}{\partial w_P}\right|}{\left|\hat{v}\cdot\frac{\partial(\vec{f}_A-\vec{f}_B)}{\partial w_P}\right|+\left|\hat{v}\cdot\frac{\partial(\vec{f}_A+\vec{f}_B)}{\partial w_P}\right|}; \atop P \in \{A,B\} \text{ and } \hat{v},\hat{w} \in \{\hat{\alpha},\hat{\beta}\} \right\}. \quad (2)$$

Each factor within this set is non-negative and never larger than unity. In most cases of interest, $\vec{f}_A$ and $\vec{f}_B$ are only functions of $\vec{r}_B - \vec{r}_A$, and, as a result, $\partial/\partial w_A$ and $-\partial/\partial w_B$ are equivalent. Therefore, $NRF_{\hat{v},w_A}^{(1)}$ and $NRF_{\hat{v},w_B}^{(1)}$ are equal and can be denoted as $NRF_{\hat{v},\hat{w}}^{(1)}$. However, we emphasize that, for $\hat{v} \neq \hat{w}$, $NRF_{\hat{v},\hat{w}}^{(1)}$ cannot be deduced from $NRF_{\hat{w},\hat{v}}^{(1)}$. Moreover, $NRF_{\hat{v},\hat{v}}^{(1)}$ cannot be inferred from $NRF_{\hat{v}}^{(0)}$, and vice versa.

## III. SYSTEM OF TWO ELECTRIC DIPOLES

Figure 2 illustrates several possible systems of two particles in free space (or any other homogenous medium). One system, that is well-known for its reciprocal interaction [26,27], is the one depicted in Fig. 2(a), where 'A' and 'B' are modeled by electric dipoles $\hat{z}\text{Re}(p_A e^{-i\omega t})$ and $\hat{z}\text{Re}(p_B e^{-i\omega t})$ with identical polarizabilities. Such a system has a parity symmetry with respect to the reversal of the $x$ axis when $y_A$ and $y_B$ are equal. As a result, $NRF_{\hat{x}}^{(0)}$ and $NRF_{\hat{x},\hat{x}}^{(1)}$ are both 'zero' irrespective of $L = x_B - x_A$. This might seem to confirm full reciprocity of the interaction. However, one should also consider the $y$ components of the forces, because $\vec{F}_A$ and $\vec{F}_B$ have $y$ components even if $\vec{F}_{A,\text{iso}}$ and $\vec{F}_{B,\text{iso}}$ do not, i.e., even if we ignore the well-known radiation pressure [28,29] originating from Ohmic loss and radiation damping. Given the parity symmetry, $NRF_{\hat{y}}^{(0)}$ and $NRF_{\hat{y},\hat{x}}^{(1)}$ are in fact 'unity' irrespective of $L$. To evaluate $NRF_{\hat{x},\hat{y}}^{(1)}$ and $NRF_{\hat{y},\hat{y}}^{(1)}$, we note that a small displacement of one particle parallel to the $y$ axis gives rise to equal and opposite phase changes in $p_A$ and $p_B$. This odd-phase symmetry leads to $NRF_{\hat{x},\hat{y}}^{(1)} = 1$ and $NRF_{\hat{y},\hat{y}}^{(1)} = 0$ irrespective of $L$.

In view of the above discussion, a non-zero $NRF_{\hat{x},\hat{x}}^{(1)}$ or $NRF_{\hat{y},\hat{y}}^{(1)}$ requires a break in the parity or the odd-phase symmetry. This happens in the case of non-identical particles with dissimilar polarizabilities [30] or when the external fields are as proposed in [31,32] and depicted in Figs. 2 (b,c). It is worth noting that a non-zero $NRF_{\hat{y},\hat{y}}^{(1)}$ is the stated goal in [31], but it is accompanied by a non-zero $NRF_{\hat{x},\hat{x}}^{(1)}$ as well. On the other hand, a non-zero $NRF_{\hat{x}}^{(0)}$ is the purpose in [30,32] but they also lead to a non-zero $NRF_{\hat{x},\hat{x}}^{(1)}$. Apart from such differences, all the non-reciprocal interparticle interactions reported so far have something in common: a certain desired value of $NRF_{\hat{x},\hat{x}}^{(1)}$ or $NRF_{\hat{y},\hat{y}}^{(1)}$ is achieved only for discrete values of $L$. In other words, the values of $NRF_{\hat{x},\hat{x}}^{(1)}$ and $NRF_{\hat{y},\hat{y}}^{(1)}$ are highly sensitive to $L$.

## IV. SYSTEM OF TWO KERKER PARTICLES

Let us now examine the conditions in which a 'one-way' interaction between two identical particles can be realized. In terms of the quantities introduced in Sec. II, such an interaction should have $NRF_{\hat{x}}^{(0)}$ and $NRF_{\hat{x},\hat{x}}^{(1)}$ equal to one-half and, furthermore, almost insensitive to $L$. We expect such an interaction to happen in the system depicted in Fig. 2(d), where the external light is a plane wave propagating parallel to $+\hat{x}$, and each particle (in the absence of the other particle) does not scatter at all in the $+\hat{x}$ direction. In such a case, we expect that 'A' feels a force $f_A$ due to the presence of 'B' while 'B' does not feel any force $f_B$ due to the presence of 'A'.

This zero forward scattering is known as the 2[nd]-Kerker condition and is perfectly met when each particle has electric and magnetic polarizabilities $\alpha_e$ and $\alpha_m = -\alpha_e/\varepsilon$, respectively, where $\varepsilon$ denotes the permittivity of the homogenous medium surrounding the particles [33,34]. Given the intuition provided in the previous paragraph, one might think that zero backward scattering, that is known as the 1[st]-Kerker condition and is met when $\alpha_m = \alpha_e/\varepsilon$, can also lead to a one-way interaction in the sense that 'B' feels a force $f_B$ due to the presence of 'A' while 'A' does not feel any force $f_A$ due to the presence of 'B'. However, as we will see later, the behavior of the interparticle forces and the non-reciprocity factors for a pair of 1[st]-Kerker particles is completely different from the one for a pair of 2[nd]-Kerker particles.

To provide an analytical solution for the interparticle forces, let us assume that the electric and magnetic fields are $y$-polarized and $z$-polarized, respectively, and therefore, treat the fields and dipoles as scalars. We also assume that the $y$ and $z$ components of $\vec{r}_A$ and $\vec{r}_B$ are small in comparison with the wavelength $\lambda_0 = 2\pi c/\omega$, where $\omega$ denotes the angular frequency of the external wave. The exerted optical force on each particle has an $x$ component only and can therefore be treated as a scalar. The force exerted on 'A' reads

$$\frac{1}{2}\text{Re}\left(p_A^* \frac{\partial E_A}{\partial x}\right) + \frac{1}{2}\mu\text{Re}\left(m_A^* \frac{\partial H_A}{\partial x}\right) - [\eta k^4/(12\pi)]\text{Re}(p_A^* m_A), \quad (3)$$

where the wavenumber k and wave impedance $\eta$ read $\omega\sqrt{\mu\varepsilon}$ and $\sqrt{\mu/\varepsilon}$, respectively, in terms of the permittivity $\varepsilon$ and permeability $\mu$ of the homogenous medium surrounding the particles [29]. The electric and magnetic dipole phasors $p_A$ and $m_A$ are related to the fields $E_A$ and $H_A$ seen by 'A' via $p_A = \alpha_e E_A(x_A)$ and $m_A = \alpha_m H_A(x_A)$. It is important to note that the derivatives in Eq. (3) are merely 'evaluated' at $x = x_A$, and are 'not' calculated and defined with respect to $x_A$ [35]. They are the derivates of the spatial field 'profiles' while $x_A$ and $x_B$ are being kept fixed. The force exerted on 'B' can be written in a similar way.

In the absence of 'B', each of the first two terms in Eq. (3) can be interpreted as the sum of a gradient force and a scattering force, where the gradient force comes from the dependence of the spatial distribution of the electromagnetic 'energy' on $x_A$, and the scattering force comes from the 'initial momenta' of the photons scattered or absorbed by 'A' [35]. Also, in the absence of 'B', the third term in Eq. (3) comes from the 'final momenta' (i.e., 'reoil') of the photons scattered by 'A' [35]. If 'A' was only an electric dipole (i.e., without a magnetic dipole), the third term would be zero because the scattering pattern of 'A' would be spatially symmetric, and the time-averaged recoil would be zero. In the presence of 'B', it is difficult (perhaps impossible) to define gradient force, scattering force, and recoil force. However, Eq. (3) is rigorous, irrespective of the entities surrounding 'A'.

The fields $E_A(x)$, $H_A(x)$, $E_B(x)$, and $H_B(x)$ can be determined for any given $x_A$ and $x_B$ by solving the following equations self-consistently:

$$\begin{cases} E_A(x) = \\ E_{ext}(x) + \alpha_e E_B(x_B)\phi(|x - x_B|) \\ \quad -\mu\alpha_m H_B(x_B)\psi(|x - x_B|), \\ \\ E_B(x) = \\ E_{ext}(x) + \alpha_e E_A(x_A)\phi(|x - x_A|) \\ \quad +\mu\alpha_m H_A(x_A)\psi(|x - x_A|) \\ \\ H_A(x) = \\ H_{ext}(x) - \alpha_e E_B(x_B)\psi(|x - x_B|) \\ \quad +\varepsilon\alpha_m H_B(x_B)\phi(|x - x_B|) \\ \\ H_B(x) = \\ H_{ext}(x) + \alpha_e E_A(x_A)\psi(|x - x_A|) \\ \quad +\varepsilon\alpha_m H_A(x_A)\phi(|x - x_A|), \end{cases} \quad (4)$$

where the propagators are given by
$$\begin{cases} 4\pi\varepsilon\, \phi(r) = (-1/r^3 + ik/r^2 + k^2/r)e^{ikr} \\ 4\pi\sqrt{\mu\varepsilon}\, \psi(r) = (ik/r^2 + k^2/r)e^{ik} \end{cases}. \quad (5)$$

For any given $x_A$ and $x_B$, Eq. (4) can be regarded as a set of four linear equations with the four unknowns $E_A(x_A)$, $H_A(x_A)$, $E_B(x_B)$, and $H_B(x_B)$. We can also find these unknowns iteratively, where the number of the iterations is in fact the number of the wave-scattering events we consider at each particle.

## V. RESULTS AND DISCUSSION

By using Eqs. (3-5) and considering one wave-scattering event at each particle, the interparticle optical forces $f_A$ and $f_B$ are calculated analytically and plotted as functions of the interparticle distance $L$ in Figs. 3(a,b) for two systems; (i) the system depicted in Fig. 2(c), where 'A' and 'B' are electric dipoles, and (ii) the system depicted in Fig. 2(d), where 'A' and 'B' are ideal 2nd-Kerker particles. As can be seen in Figs. 3(a,b), the oscillations of $f_A(L)$ for the pair of 2nd-Kerker particles are somewhat like those for the pair of electric dipoles. This is not accidental and is rooted in the fact that the propagators $4\pi\varepsilon\phi(r)$ and $4\pi\sqrt{\mu\varepsilon}\psi(r)$ in Eq. (5) are equal if one ignores the quasi-static term $-e^{ikr}/r^3$ in $4\pi\varepsilon\phi(r)$.

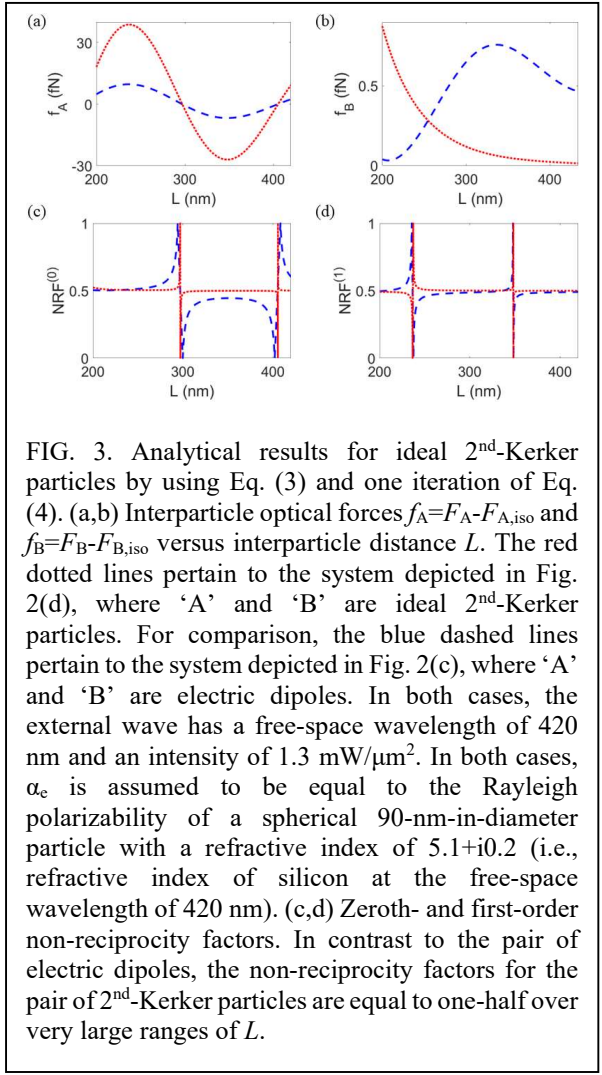

FIG. 3. Analytical results for ideal 2nd-Kerker particles by using Eq. (3) and one iteration of Eq. (4). (a,b) Interparticle optical forces $f_A=F_A-F_{A,iso}$ and $f_B=F_B-F_{B,iso}$ versus interparticle distance $L$. The red dotted lines pertain to the system depicted in Fig. 2(d), where 'A' and 'B' are ideal 2nd-Kerker particles. For comparison, the blue dashed lines pertain to the system depicted in Fig. 2(c), where 'A' and 'B' are electric dipoles. In both cases, the external wave has a free-space wavelength of 420 nm and an intensity of 1.3 mW/μm$^2$. In both cases, $\alpha_e$ is assumed to be equal to the Rayleigh polarizability of a spherical 90-nm-in-diameter particle with a refractive index of 5.1+i0.2 (i.e., refractive index of silicon at the free-space wavelength of 420 nm). (c,d) Zeroth- and first-order non-reciprocity factors. In contrast to the pair of electric dipoles, the non-reciprocity factors for the pair of 2nd-Kerker particles are equal to one-half over very large ranges of $L$.

As can be seen in Figs. 3(c,d), for the pair of electric dipoles, the non-reciprocity factors are quite sensitive to $L$, and equal to one-half only for certain values of $L$. In contrast, for the pair of ideal 2nd-Kerker particles, the non-reciprocity factors are equal to one-half over very large ranges of $L$, which confirms a perfect one-way interaction. The jumps and dips observed in $NRF_{\hat{x}}^{(0)}$ in Fig. 3(c) take place at the values of $L$ for which $f_A$ and $f_B$ are both very small. The jumps and dips observed in $NRF_{\hat{x},\hat{x}}^{(1)}$ in Fig. 3(d) take place at the values of $L$ for which $\partial f_A/\partial L$ and $\partial f_B/\partial L$ are both very small. For the pair of ideal 2nd-Kerker particles, all jumps and dips in the non-reciprocity factors happen within extremely narrow regions.

Let us now examine the 2nd-Kerker condition in more detail. Due to different loss mechanisms, including radiation damping that is always present, the imaginary parts of $\alpha_e$ and $\alpha_m$ are both positive in the absence of optical gain [28]. Therefore, the condition $\alpha_m = -\alpha_e/\varepsilon$ of zero forward scattering cannot be perfectly met in the absence of optical gain. From another viewpoint, the optical theorem, that is a manifestation of causality, prohibits zero forward scattering. Nevertheless, forward scattering can still be very small for gainless dielectric particles, e.g., for silicon particles of suitable size [36]. By using the Mie theory and considering the frequency-dependent refractive index of silicon, we found that the ratio of forward to backward scattering is small for a 90-nm-in-diameter silicon particle at a free-space wavelength of 420 nm.

By using the finite element method and calculating the Maxwell stress tensor, the interparticle optical forces $f_A$ and $f_B$ are calculated rigorously and plotted as functions of the center-to-center distance $L$ in Figs. 4(a,b) for the system depicted in Fig. 2(d), where 'A' and 'B' are now 90-nm-in-diameter silicon particles, and the external wave has a free-space wavelength of 420 nm. As can be seen in Figs. 4(c,d), the non-reciprocity factors are almost equal to one-half over large ranges of $L$. This shows that a nearly-one-way interaction can be realized by using dielectric particles without incorporating optical gain. The small noise observed in $NRF_{\hat{x},\hat{x}}^{(1)}(L)$ in Fig. 4(d) is due to numerical errors in calculating the derivatives $\partial f_A/\partial L$ and $\partial f_B/\partial L$.

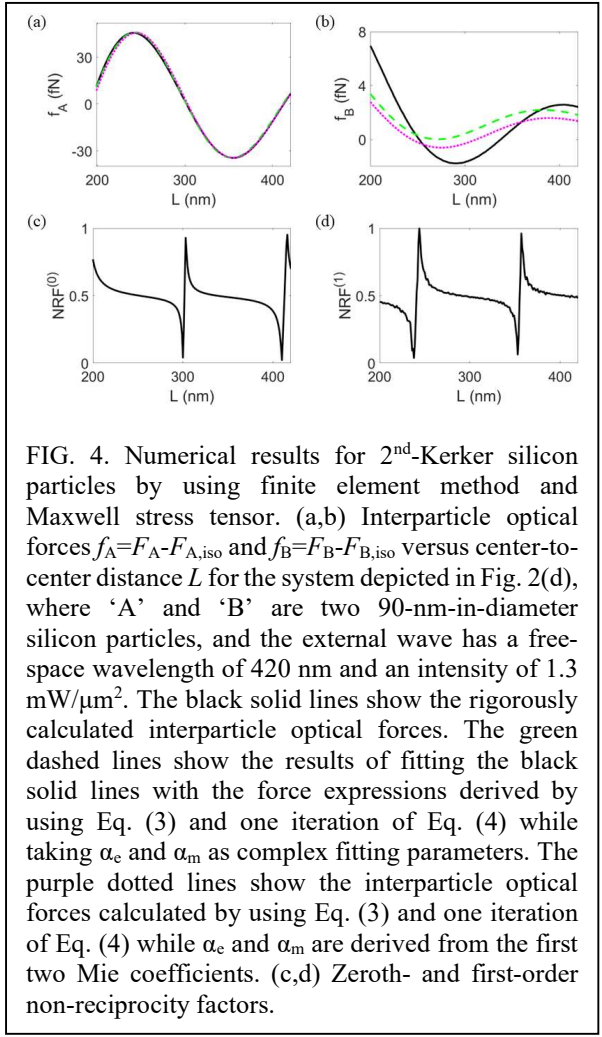

FIG. 4. Numerical results for 2$^{nd}$-Kerker silicon particles by using finite element method and Maxwell stress tensor. (a,b) Interparticle optical forces $f_A = F_A - F_{A,iso}$ and $f_B = F_B - F_{B,iso}$ versus center-to-center distance $L$ for the system depicted in Fig. 2(d), where 'A' and 'B' are two 90-nm-in-diameter silicon particles, and the external wave has a free-space wavelength of 420 nm and an intensity of 1.3 mW/µm$^2$. The black solid lines show the rigorously calculated interparticle optical forces. The green dashed lines show the results of fitting the black solid lines with the force expressions derived by using Eq. (3) and one iteration of Eq. (4) while taking $\alpha_e$ and $\alpha_m$ as complex fitting parameters. The purple dotted lines show the interparticle optical forces calculated by using Eq. (3) and one iteration of Eq. (4) while $\alpha_e$ and $\alpha_m$ are derived from the first two Mie coefficients. (c,d) Zeroth- and first-order non-reciprocity factors.

One might like to see whether a 90-nm-in-diameter silicon particle at a free-space wavelength of 420 nm can be modeled as the combination of an electric dipole and a magnetic dipole. To this end, one way is to examine the Mie coefficients $a_1$ and $b_1$ corresponding to the electric and magnetic dipoles and show that they are orders of magnitude larger than all other Mie coefficients. The other way is to fit $f_A(L)$ and $f_B(L)$, that are calculated rigorously using the finite element method and the Maxwell stress tensor, with the expressions derived by using Eqs. (3-5) with $\alpha_e/\varepsilon$ and $\alpha_m$ as the fitting parameters. The results of both procedures are shown in Figs. 4(a,b).

It should be emphasized that the particle size of 90 nm and the free-space wavelength of 420 nm are merely the values that yield a small forward scattering for a silicon particle. In other words, we did not do a rigorous optimization to find the particle size and the free-space wavelength to achieve the least possible forward scattering or the closest possible non-reciprocity factors to one-half. If one does such an optimization, the resulting interaction becomes closer to a perfect one-way interaction, and the non-reciprocity factors become more similar to the red dotted lines in Figs. 3(c,d).

Let us now answer the question of why we cannot have a one-way interaction based on a pair of 1$^{st}$-Kerker particles, i.e., particles with zero backward scattering. For a system of two 1$^{st}$-Kerker particles, like the one depicted in Fig. 5(a), one might think that 'B' feels a force $f_B$ due to the presence of 'A' while 'A' does not feel any force $f_A$ due to the presence of 'B'. By using Eqs. (3-5) and considering one wave-scattering event at each particle, the interparticle optical forces $f_A$ and $f_B$ are calculated analytically and plotted as functions of the interparticle distance $L$ in Fig. 5(b) for the system depicted in Fig. 5(a), where 'A' and 'B' are ideal 1$^{st}$-Kerker particles.

As can be seen in Fig. 5(b), the behavior of the interparticle forces for a pair of 1$^{st}$-Kerker particles is completely different from the one plotted in Figs. 3(a,b) for a pair of 2$^{nd}$-Kerker particles. The binding force $f_A - f_B$ is mostly positive (i.e., attractive) and have two unstable zeros for the pair of 1$^{st}$-Kerker particles, whereas it oscillates and have both unstable and stable zeros for the pair of 2$^{nd}$-Kerker particles. Moreover, unlike the case of the pair of 2$^{nd}$-Kerker particles, $f_A$ and $f_B$ are of the same order of magnitude for the pair of 1$^{st}$-Kerker particles. This is why the non-reciprocity factors plotted in Figs. 5(c,d) do not show any signature of a one-way interaction for the pair of 1st-Kerker particles.

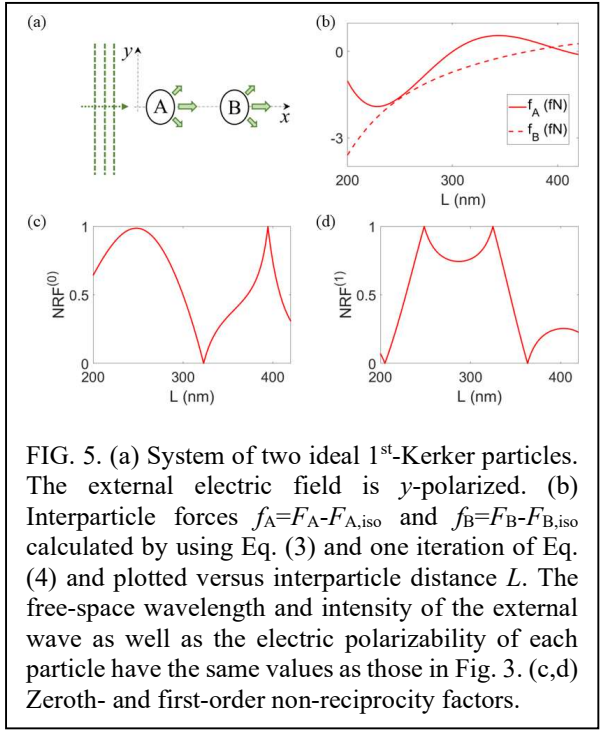

FIG. 5. (a) System of two ideal 1st-Kerker particles. The external electric field is $y$-polarized. (b) Interparticle forces $f_A = F_A - F_{A,iso}$ and $f_B = F_B - F_{B,iso}$ calculated by using Eq. (3) and one iteration of Eq. (4) and plotted versus interparticle distance $L$. The free-space wavelength and intensity of the external wave as well as the electric polarizability of each particle have the same values as those in Fig. 3. (c,d) Zeroth- and first-order non-reciprocity factors.

The resented results for the pair of 1st-Kerker particles clearly show that the behavior of the interparticle forces cannot be deduced directly from the scattering pattern of the individual particles. The sole reason that the pair of ideal 2nd-Kerker particles yields a perfect one-way interaction is not that each particle has zero forward scattering. The one-way interaction is the result of the π-phase difference between $\alpha_e$ and $\alpha_m$ for each of the two particles. The zero forward scattering itself is another 'result' of that phase relationship.

## VI. CONCLUDING REMARKS

In conclusion, we have shown that a one-way optomechanical interaction can be established between two nanoparticles 'A' and 'B' in the sense that the exerted optical force on 'A' due to the presence of 'B' is considerable and highly dependent on the interparticle distance while the exerted optical force on 'B' due to the presence of 'A' is zero or negligible.

As the interaction is mediated by optical fields, it can be controlled externally. The one-way interaction proposed in this article is based on particles with directional scattering patterns. However, we elucidated that scattering patterns can be very misleading about the interparticle forces.

To precisely quantify the level of non-reciprocity in any optomechanical interaction between two particles, including the one-way interaction proposed in this article as well as the interactions reported so far, we introduced the zeroth- and first-order non-reciprocity factors. These factors are functions of the interparticle distance. For a general three-dimensional problem, we have three zeroth-order factors and nine first-order factors.

The interparticle interaction we discussed is scalable to systems comprising many particles. In complex out-of-equilibrium systems, non-reciprocal interactions not only lead to mesoscopic thermodynamic effects such as heat pumping and refrigeration [20,21], but they can also determine the anomalous diffusion of constituents [37,38]. The non-reciprocal interactions may also assist in creating exotic phases of active matter [13,14,39,40]. Furthermore, in view of recent experimental advancements in building arrays of levitated nanoparticles [41], the one-way interaction we proposed could constitute a building block for the simulation of non-Hermitian quantum models [42,43].

[1] A. O. Caldeira and A. J. Leggett, "Influence of Dissipation on Quantum Tunneling in Macroscopic Systems," Phys. Rev. Lett. 46, 211 (1981).
[2] D. F. Walls and G. J. Milburn, *Quantum Optics*, (Springer, Berlin, 2008).
[3] A. A. Clerk, M. H. Devoret, S. M. Girvin, F. Marquardt, and R. J. Schoelkopf, "Introduction to quantum noise, measurement, and amplification," Rev. Mod. Phys. 82, 1155 (2010).
[4] J. Millen, T. Deesuwan, P. Barker, and J. Anders, "Nanoscale temperature measurements using non-equilibrium Brownian dynamics of a levitated nanosphere," Nat. Nanotechnol. 9, 425 (2014).

[5] F. Marquardt, J. P. Chen, A. A. Clerk, and S. M. Girvin, "Quantum Theory of Cavity-Assisted Sideband Cooling of Mechanical Motion," Phys. Rev. Lett 99, 093902 (2007).

[6] J. D. Teufel et al., "Sideband cooling of micromechanical motion to the quantum ground state," Nature 475, 359 (2011).

[7] U. Delić, M. Reisenbauer, D. Grass, N. Kiesel, V. Vuletić, and M. Aspelmeyer, "Cavity Cooling of a Levitated Nanosphere by Coherent Scattering," Phys. Rev. Lett. 122, 123602 (2019).

[8] D. Windey, C. Gonzalez-Ballestero, P. Maurer, L. Novotny, O. Romero-Isart, and R. Reimann, "Cavity-Based 3D Cooling of a Levitated Nanoparticle via Coherent Scattering," Phys. Rev. Lett. 122, 123601 (2019).

[9] U. Delić et al., "Cooling of a levitated nanoparticle to the motional quantum ground state," Science 367, 892 (2020).

[10] H. B. Callen and T. A. Welton, "Irreversibility and Generalized Noise," Phys. Rev. Lett. 83, 34 (1951).

[11] I. Söllner et al., "Deterministic photon–emitter coupling in chiral photonic circuits," Nat. Nanotechnol. 10, 775 (2015).

[12] P. Lodahl et al., "Chiral quantum optics," Nature 541, 473 (2017).

[13] M. Fruchart, R. Hanai, P. B. Littlewood, and V. Vitelli, "Non-reciprocal phase transitions," Nature 592, 363 (2021).

[14] M. J. Bowick, N. Fakhri, M. C. Marchetti, and S. Ramaswamy, "Symmetry, Thermodynamics, and Topology in Active Matter," Phys. Rev. X 12, 010501 (2022).

[15] S. Osat and R. Golestanian, "Non-reciprocal multifarious self-organization," Nat. Nanotechnol. 18, 79 (2023).

[16] S. H. L. Klapp, "Non-reciprocal interaction for living matter," Nat. Nanotechnol. 18, 8 (2023).

[17] T. Helbig et al., "Generalized bulk–boundary correspondence in non-Hermitian topolectrical circuits," Nat. Phys. 16, 747 (2020).

[18] M. Brandenbourger, X. Locsin, E. Lerner, and C. Coulais, "Non-reciprocal robotic metamaterials," Nat. Commun. 10, 4608 (2019).

[19] H. Xu, L. Jiang, A. A. Clerk, and J. G. E. Harris, "Nonreciprocal control and cooling of phonon modes in an optomechanical system," Nature 568, 65 (2019).

[20] S. A. M Loos and and S. H. L. Klapp, "Irreversibility, heat and information flows induced by non-reciprocal interactions," New J. Phys. 22, 123051 (2020).

[21] S. A. M Loos, S. Arabha, A. Rajabpour, A. Hassanali, and É. Roldán, "Nonreciprocal forces enable cold-to-hot heat transfer between nanoparticles," Sci. Rep. 13, 4517 (2023).

[22] K. Chen, P. Santhanam, S. Sandhu, L. Zhu, and S. Fan, "Heat-flux control and solid-state cooling by regulating chemical potential of photons in near-field electromagnetic heat transfer," Phys. Rev. B 91, 134301 (2015).

[23] L. Zhu, A. Fiorino, D. Thompson, R. Mittapally, E. Meyhofer, and P. Reddy, "Near-field photonic cooling through control of the chemical potential of photons," Nature 566, 239 (2019).

[24] S. Buddhiraju, W. Li, and S. Fan, "Photonic refrigeration from time-modulated thermal emission," Phys. Rev. Lett. 124, 077402 (2020).

[25] J. R. Deop-Ruano and A. Manjavacas, "Control of the radiative heat transfer in a pair of rotating nanostructures," Phys. Rev. Lett. 130, 133605 (2023).

[26] K. Dholakia and P. Zemánek, "Colloquium: Gripped by light: Optical binding," Rev. Mod. Phys. 82, 1767 (2010).

[27] Y. Arita, G. D. Bruce, E. M. Wright, S. H. Simpson, P. Zemánek, and K. Dholakia, "All-optical sub-Kelvin sympathetic cooling of a levitated microsphere in vacuum," Optica 9, 1000 (2022).

[28] B. T. Draine, "The Discrete-Dipole Approximation and Its Application to Interstellar Graphite Grains," Astrophys. J. 333, 848 (1988).

[29] P. C. Chaumet and A. Rahmani, "Electromagnetic force and torque on magnetic and negative-index scatterers," Opt. Exp. 17, 2224 (2009).

[30] S. Sukhov, A. Shalin, D. Haefner, and A. Dogariu, "Actio et reactio in optical binding," Opt. Exp. 23, 247 (2015).

[31] J. Rieser et al., "Tunable light-induced dipole-dipole interaction between optically levitated nanoparticles," Science 377, 987 (2022).

[32] C. W. Peterson, J. Parker, S. A. Rice, and N. F. Scherer, "Controlling the Dynamics and Optical Binding of Nanoparticle Homodimers with Transverse Phase Gradients," Nano Lett. 19, 897 (2019).

[33] M. Kerker, D. S. Wang, and C. L. Giles, "Electromagnetic scattering by magnetic spheres," J. Opt. Soc. Am. 73, 765 (1983).

[34] J. Olmos-Trigo et al., "Kerker Conditions upon Lossless, Absorption, and Optical Gain Regimes," Phys. Rev. Lett. 125, 073205 (2020).

[35] A. M. Jazayeri, "Critical ambient pressure and critical cooling rate in optomechanics of electromagnetically levitated nanoparticles," JOSA B 38, 3652 (2021).

[36] Y. H. Fu, A. I. Kuznetsov, A. E. Miroshnichenko, Y. F. Yu, and B. Luk'yanchuk, "Directional visible light scattering by silicon nanoparticles," Nat. Commun. 4, 1527 (2013).

[37] K. M. Douglass, S. Sukhov, and A. Dogariu, "Superdiffusion in optically controlled active media," Nat. Phot. 6, 834 (2012).

[38] E. Arvedson, M. Wilkinson, B. Mehlig, and K. Nakamura, "Staggered ladder spectra," Phys. Rev. Lett. 96, 030601 (2006).

[39] F. Jendrzejewski, *et al.*, "Three-dimensional localization of ultracold atoms in an optical disordered potential," Nat. Phys. 8, 398 (2012).

[40] S. Shankar, A. Souslov, M. J. Bowick, M. C. Marchetti, and V. Vitelli, "Topological active matter," Nat. Rev. Phys. 4, 380 (2022).

[41] J. Vijayan, *et al.*, "Scalable all-optical cold damping of levitated nanoparticles," Nat. Nanotechnol. 18, 49 (2023).

[42] Q. Liang, *et al.*, "Dynamic signatures of non-Hermitian skin effect and topology in ultracold atoms," Phys. Rev. Letters 129, 070401 (2022).

[43] A. J. Daley, *et al.*, "Practical quantum advantage in quantum simulation," Nature 607, 667 (2022).